\documentclass[reprint,superscriptaddress,amsmath,amssymb,aps,pre]{revtex4-1}

\usepackage{graphicx}
\usepackage{graphics}
\usepackage{amsmath}
\usepackage{dcolumn}
\usepackage{bm}
\usepackage{subfig}

\begin{document}

\title{Superior ionic and electronic properties of ReN$_2$ monolayers as Na-ion battery electrodes}
\author{Shi-Hao Zhang}
\affiliation{Beijing National Laboratory for Condensed Matter Physics, Institute of Physics, Chinese Academy of Sciences, Beijing 100190, China.}
\author{Bang-Gui Liu}%
 \email{bgliu@iphy.ac.cn}
\affiliation{Beijing National Laboratory for Condensed Matter Physics, Institute of Physics, Chinese Academy of Sciences, Beijing 100190, China.}
\affiliation{School of Physical Sciences, University of Chinese Academy of Sciences, Beijing 100190, China}

\date{\today}

\begin{abstract}
Excellent two-dimensional electrode materials can be used to design high-performance alkali-metal-ion batteries. Here, we propose ReN$_2$ monolayer as a superior two-dimensional material for sodium-ion batteries. Total-energy optimization results in a buckled tetragonal structure for ReN$_2$ monolayer, and our phonon spectrum and elastic moduli prove its dynamical and mechanical stability. Further investigation shows that it is metallic and still keep metallic feature after the adsorption of Na or K atoms, its lattice parameter changes by only 3.2\% or 3.8\% after absorption of Na or K atoms. Our study shows that its maximum capacity reaches 751 mA h/g for Na-ion batteries or 250 mA h/g for K-ion batteries, and its diffusion barrier is only 0.027 eV for Na atom or 0.127 eV for K atom. The small lattice change, high storage capacity, metallic feature, and extremely low ion diffusion barriers make the ReN$_2$ monolayer become superior electrode materials for Na-ion rechargeable batteries with ultrafast charging/discharging processes.
\end{abstract}

\pacs{Valid PACS appear here}
\maketitle

\section{Introduction}

Recent years have witnessed a booming development in two-dimensional materials since the experimental discovery of graphene~\cite{1,2,3,4}. The two-dimensional materials are promising for applications in many fields, such as field-effect transistors~\cite{5,6}, phototransistors~\cite{7,8}, p-n junctions~\cite{9,10}, supercapacitors~\cite{11,12}, and batteries~\cite{13,14}. Besides graphene, two-dimensional materials such as phosphorene and transition-metal dichalcogenides (TMDC) exhibit many novel physical properties which do not exist in their bulk counterparts.
Since the commercialization of the lithium-ion batteries in 1991, rechargeable lithium-ion batteries, with the reversible capacity and cycle life, have attracted huge interest~\cite{l1,l2,l3,l4,l6,l7,l8,l9}. Nevertheless, the Li-ion batteries still need to be improved for better reversibility and ion diffusivity and smaller volume change ~\cite{s1,s2,s3}. On the other hand, it is highly desirable to develop new anode materials for high-performance ion batteries. Because of the large natural abundance of Na (23000 ppm) and K (17000 ppm) in comparison to that of Li (20 ppm) in the earth's crust~\cite{s4}, Na-ion and K-ion batteries have attracted much attention. To date, a tremendous number of 2D materials, including graphene systems~\cite{s6,s7}, transition-metal dichalcogenides~\cite{s8,s9,s10}, transition-metal carbides~\cite{s11,s12,s13,added}, and metal nitrides~\cite{s14}, have been studied because of their excellent electrochemical performance as battery anode materials. The storage capacity for most of the two-dimensional materials is between 200 and 600 mA h g$^{-1}$ and the ion diffusion barrier ranges from 0.1 to 0.6 eV. It is still necessary to seek new materials for ion batteries with high capacity and low ion diffusion barrier.

Here, through first-principles calculation we study ReN$_2$ monolayers as two-dimensional structures and investigate their electronic and mechanical properties for superior anode materials of alkali-metal-ion batteries. We obtain the stable tetragonal structure of ReN$_2$ monolayer in terms of our calculated phonon spectra and in-plane stiffness constants. According to elastic theory, the tetragonal ReN$_2$ can be freestanding without the support of substrate. Our calculations with the PBE functional and HSE functional with/without the spin-orbit coupling show that the ReN$_2$ monolayer is metallic. After the absorption of Na and K atoms, the system still keep the metallic feature which is advantageous for the applications in Na-ion and K-ion batteries. The storage capacity of the ReN$_2$ monolayer is 751 mA h g$^{-1}$ or 250 mA h g$^{-1}$ if taken as Na-ion and K-ion anode materials, and the corresponding ion diffusion barrier, 0.027 eV for Na or 0.127 eV for K, is very small. The high capacity and the extremely low diffusion barrier for Na atom make the ReN$_2$ monolayer a superior anode material. More detailed results will be presented in the following.

\section{Computational methods}

The first-principles calculations are done with the projector-augmented wave (PAW) potential method~\cite{22} as implemented in the Vienna ab initio simulation package software (VASP)~\cite{23}. We take the generalized gradient approximation (GGA), accomplished by Perdew, Burke, and Ernzerhof (PBE)~\cite{24}, for the exchange-correlation functional. The kinetic energy cutoff of the plane waves is set to 600 eV. For both optimization and static calculation, the Brillouin zone integration is carried out with a 10$\times$10$\times$1 special $\Gamma$-centered k-point mesh following the convention of Monkhorst-Pack~\cite{25}. All atomic positions are fully optimized with the conjugate gradient optimization until all the Hellmann-Feynman forces on each atom are less than 0.01 eV/\AA{} and the total energy difference between two successive steps is smaller than 10$^{-6}$ eV. Furthermore, phonon dispersion calculation in terms of the density functional perturbation theory, by using the PHONOPY program~\cite{29}, is performed to ensure the structural stability of the monolayers. We take the 4$\times$4$\times$1 supercell for calculating the phonon spectra of the 2D structures. In order to make further confirmation, band dispersion calculations with Heyd-Scuseria-Ernzerhof (HSE) hybrid functional~\cite{26,27,28} are carried out, with the mixing rate of the HF exchange potential being 0.25. The semi-empirical correction scheme of Grimme (DFT-D2)~\cite{D2} is employed to evaluate the effect of van der Waals (vdW) interactions on Na/K ion adsorption. In the calculation of Na/K diffusion, we use nudged elastic band (NEB) method to get the ion diffusion barrier.


\section{Results and discussion}

\subsection{Structures and stability}

The three-dimensional materials of rhenium dinitride ReN$_{2}$ have been synthesized by metathesis reaction under high pressure~\cite{19}, and X-ray Diffraction(XRD) shows that the samples have the same structure as the three-dimensional MoS$_{2}$-like hexagonal structure. A systematical first-principles investigation shows that the three-dimensional MoN$_2$-like structure is dynamically stable for ReN$_{2}$ and there is a three-dimensional tetragonal structure with lower total energy~\cite{20,21}. In addition, experiment indicates that the three-dimensional ReN$_2$ may be layered in terms of its compressibility~\cite{19}.

\begin{figure}[!htbp]
\includegraphics[width=0.48\textwidth]{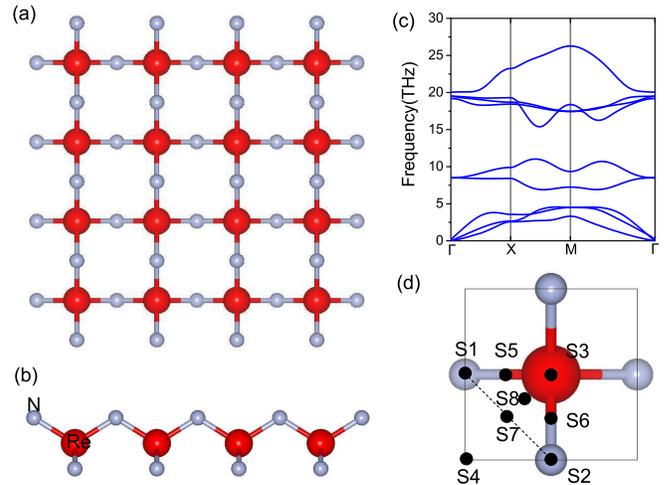}
\caption{\label{fig1}The top (a) and side (b) views of the 2D tetragonal structure, the phonon spectra (c), and the possible sites for Na/K absorption (d) of the ReN$_2$ monolayer.}
\end{figure}

 As for the two-dimensional structure of ReN$_2$ monolayer, four possible configurations are considered: T-phase structure, T$^{\prime}$-phase structure, H-phase structure, and tetragonal structure. The T-phase structure, which has been found in other two-dimensional materials~\cite{30}, is proved to be unstable for ReN$_2$ in terms of its phonon spectrum result. Distorted T$^{\prime}$-phase structure~\cite{31,32,33,34} is also proved to be impossible because our phonon spectrum calculation show that there are very large negative phonon frequencies near the $\Gamma$ point. Both H-phase structure and tetragonal structure have the phonon spectra without imaginary phonon modes. The total energy of the tetragonal structure is lower than the hexagonal one by 0.33 eV per formula unit. Therefore, the two-dimensional tetragonal structure is stable. It (P$\bar{4}$m2) is made up with three atom planes as shown in FIG.~\ref{fig1}. The comparative study of spin-polarized and spin-unpolarized  calculations show that the tetragonal structure is non-magnetic. Its in-plane lattice parameters $a$ is 3.178 \AA{}, and the vertical distance between the top and bottom nitrogen planes is 1.96 \AA{}. The Re-N bond length $l_{\rm Re-N}$ is 1.87 \AA{}. The cohesive energy per formula unit is defined as $E_{coh}=E_{\rm Re}+E_{\rm N_2}-E_{\rm ReN_2}$, where $E_{\rm Re}$ and $E_{\rm N_2}$ are the total energies of isolated Re atom and N$_2$ molecule. The cohesive energies 6.44 eV is much larger than 1.87 eV for the case of MoN$_2$ \cite{mon2ce}, which reveals that the synthesis of the buckled ReN$_2$ monolayer is accomplishable.

In order to analyze the dynamical stability of the 2D tetragonal structure, the phonon spectrum calculations are performed by using the PHONOPY program~\cite{29}, and the calculated results are presented in FIG.~\ref{fig1}. It is clear that there are nine phonon branches including three acoustic branches and six optical phonon bands. Non-existence of negative phonon frequencies in FIG.~\ref{fig1} proves the dynamical stability of the 2D tetragonal structures. It can be seen in FIG.~\ref{fig1} that in the vicinity of the $\Gamma$ point, the acoustical branches show linear dispersion, and the out-of-plane acoustical branch appears to be softer than the other two due to the special mode in the two-dimensional materials~\cite{35}.

To inspect the mechanical stability, we calculate the elastic constants of the ReN$_2$ monolayer: C$_{11}$ = 107.4 N/m, C$_{22}$ = 108.0 N/m, C$_{12}$ = 74.3 N/m, and C$_{66}$ = 218.8 N/m. The Young's modulus $Y$ is equivalent to 56.3 N/m. For two-dimensional materials, only $C_{11}$, $C_{12}$, $C_{16}$, $C_{26}$, $C_{66}$ and $C_{22}$ are meaningful quantities, and the criteria of mechanical stability require that $C_{11}C_{22} > C_{12}C_{21}$ and $C_{66}>0$~\cite{prb} are satisfied. Thus the tetragonal ReN$_2$ monolayer has the mechanical stability. Assuming that we have the square flake with the edge length $l$, the ratio between the out-of plane deformation $h$ induced by its own gravity and the edge length $l$ is $h/l \approx (\rho gl/Y)^{1/3}$, where g being the gravitational acceleration and $\rho$ the density of the two-dimensional material. Here, the density of the ReN$_2$ monolayer is $3.54 \times 10^{-6}$ kg$\cdot$m$^{-2}$, and then we can obtain $h/l \approx (0.62l \times 10^{-12})^{1/3}$ for the ReN$_2$monolayer, where $l$ is in micrometer. Even for $10^{4}$ $\mu$m$^2$ flakes, the ratio is only $3.96 \times 10^{-4}$, which is smaller than the previous result of Ca$_2$N freestanding monolayer, $6.31\times 10^{-4}$~\cite{36}. Therefore, the 2D ReN$_2$ monolayer can keep its stability without the support of a substrate.

\begin{figure}[!htbp]
\includegraphics[width=0.5\textwidth]{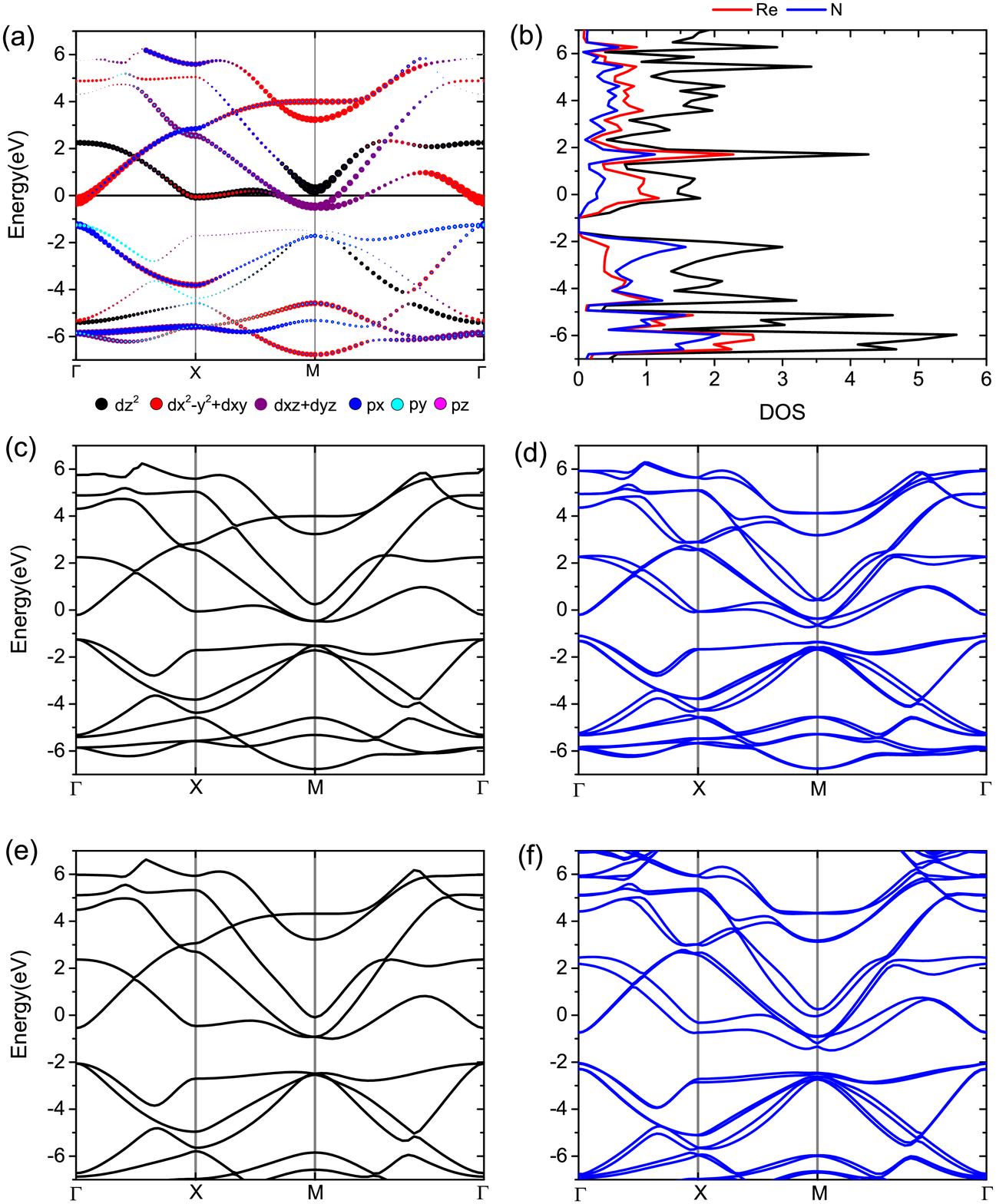}
\caption{\label{fig2}The energy bands (a) and density of states (b) of the ReN$_2$ monolayer, the PBE energy bands without SOC (c) and with SOC (d), and the HSE energy bands without SOC (e) and with SOC (f).}
\end{figure}

\subsection{Electronic structures}

The 2D tetragonal structure obeys the $S_4$ symmetry group. For the electronic wave function of the d orbits, the parity of mirror inversion is even for $d_{z^2}$, $d_{xy}$, and $d_{x^2-y^2}$, and odd for $d_{xz}$ and $d_{yz}$. The band dispersions of the 2D materials without the spin-orbit coupling, with the weights of atomic orbitals indicated, are presented in FIG.~\ref{fig2}. The size of symbol is proportional to the weight of the orbital. In the energy spectra, all the N p bands are filled, and the nearly empty bands come from Re d states. Near the $\Gamma$ point, the band near the Fermi level is mainly from Re d$_{x^2-y^2}$, and near the X point, the band near the Fermi level consists mainly of Re d$_{z^2}$ and d$_{x^2-y^2}$.

The band structure shows that the two-dimensional ReN$_2$ monolayer is metallic, but we need to confirm that the system keep metallic feature with HSE calculation or spin-orbit coupling (SOC), because PBE calculation always underestimates the semiconductor gaps and the spin-orbit coupling can open a gap in the transition-metal system. For this purpose, we  calculate the energy spectra with HSE with SOC, and the calculated results are presented in FIG.~\ref{fig2}. As FIG.~\ref{fig2} shows, the tetragonal ReN$_2$ monolayer is still metallic. This is advantageous for its applications in Na-ion and K-ion battery technology.

\subsection{Ion adsorption}

A 2$\times$2 supercell of the ReN$_2$ monolayer is used as the substrate for the adsorption of the Na atom and K atom. After adsorbing a metal atom, the chemical formula of the system can be written as ARe$_4$N$_8$, where A represents Na atom or K atom. The adsorption energy $E_{a}$ is defined by $E_{a}=E_{\rm ARe_4N_8}-E_{\rm A}-E_{\rm Re_4N_8}$ where $E_{\rm A}$ is the total energy of bulk A metal per atom, and $E_{\rm ARe_4N_8}$ and $E_{\rm Re_4N_8}$ are the total energy of the Re$_4$N$_8$ monolayer with and without A adsorption. According to the symmetry of the Re$_4$N$_8$ monolayer, eight possible sites are considered, as shown in FIG.~\ref{fig1}(d). Four of the eight sites remain after our geometrical optimization, and their adsorption energies are summarized in TABLE.~\ref{table1}. We can see that the adsorption energies of all the four sites are negative, which means that Na/K atom prefers to be adsorbed on the host material instead of forming a cluster.

The unit cell of the ReN$_2$ monolayer includes one Re atom, one high-position N atom, and one low-position N atom. According to the adsorption energy, Na/K atom prefers to stay at the position above the low-position N atom (S2 site), and the distance between Na atom and the nitrogen atom right below it is 3.49 \AA{} (3.89 \AA{} for K atom). The adsorption energy at the hollow site (S4) is slightly larger than the adsorption energy at the S2, and much smaller than the adsorption energies at the S1 and S3 sites. Because of the Coulomb repulsion between Na/K atom and Re atom, the absorbed Na/K atom will not stay at the S3 site. Nearest the S1 and S4 sites are the high-position N atoms, but nearest the S2 site are two Re atoms. Thus the absorption of Na/K atom at the S2 site is lowest among the four possible sites, which can explain the absorption behavior of Na/K atom.

\begin{figure}[!htbp]
\includegraphics[width=0.48\textwidth]{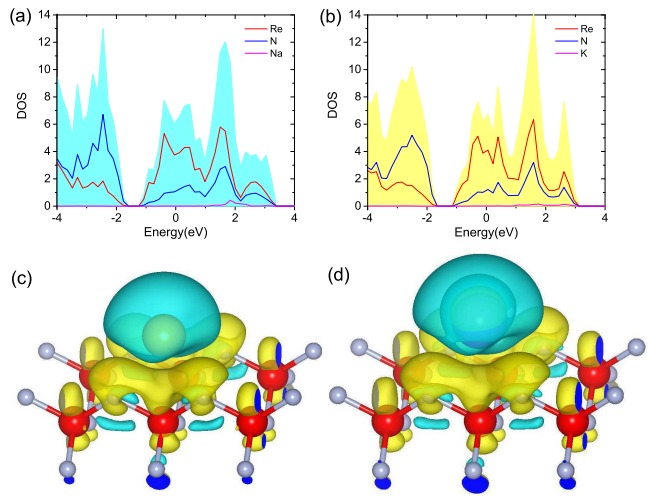}
\caption{\label{fig3}The density of states (DOS) of NaRe$_4$N$_8$ (a) and KRe$_4$N$_8$ (b). The charge density difference $\Delta \rho$ with the absorption of Na atom (c) and K atom (d) with the isosurface level of 0.001 e/\AA{}$^3$. The yellow area represents the positive charge density difference and the blue area represents the negative one.}
\end{figure}

To further understand the absorption of Na/K atom, we make the Bader charge analysis and the results are summarized in TABLE.~\ref{table1}. The existence of charge transference by Na/K atom reveals that the adsorption is chemical, which can be regarded as redox reaction during the battery operation. The calculated charge density difference is showed in FIG.~\ref{fig3}, which is defined by $\Delta \rho = \rho (\rm ARe_4N_8)-\rho (\rm A)-\rho (\rm Re_4N_8)$. This confirms the existence of chemical adsorption. The larger atom (K atom) reduces the adsorption energy difference between different sites. We also calculate the density of states (DOS) of the Re$_4$N$_8$ after absorbing Na/K atom. The calculated results show that the system still keep metallic character, which is advantageous for making electrode materials from the ReN$_2$ monolayer.

\begin{table}[!htbp]
\caption{\label{table1}The site-dependent adsorption energies of Na atom ($E_{a{\rm Na}}$) and K atom ($E_{a{\rm K}}$), and the charge transferred by Na atom ($q_{\rm Na}$) and K atom ($q_{\rm K}$).}
\begin{ruledtabular}
\begin{tabular}{ccccc}
Sites & $E_{a{\rm Na}}$ (eV) & $q_{\rm Na}$ (e) & $E_{a{\rm K}}$ (eV) & $q_{\rm K}$ (e)\\ \hline
S1   & -0.97 & 0.88 & -1.46 & 0.88 \\
S2   & -1.80 & 0.87 & -2.12 & 0.88 \\
S3   & -1.31 & 0.88 & -1.70 & 0.88 \\
S4   & -1.78 & 0.84 & -1.99 & 0.88 \\
\end{tabular}
\end{ruledtabular}
\end{table}

\subsection{Key storage parameters}

Open circuit voltage (OCV) and theoretical storage capacity are the important parameters to describe the performance of electrode materials. The charge/discharge process of ReN$_2$ monolayer can be described by ReN$_2$ + $x$A$^+$ + $xe^- \leftrightarrow$ A$_x$ReN$_2$. For this reaction, the average open circuit voltage can be defined by $V_{ave} = (E_{\rm ReN_2} + xE_{\rm A} - E_{{\rm A}_x{\rm ReN_2}})/xe$ when we ignore the volume and entropy effects during the alkali metal adsorption process. $E_{\rm ReN_2}$ and $E_{{\rm A}_x{\rm ReN_2}}$ are the total energies of the ReN$_2$ monolayer before and after the adsorption of A (Na or K) atom. In order to calculate the storage capacity, we choose 35\AA{} as the thickness of vacuum slab to avoid the interaction between neighboring layers. In the process of Na/K intercalation, we calculate the average adsorption energy layer by layer, which is defined by $E_n = (E_{{\rm A}_{8n}{\rm Re_4N_8}}-E_{{\rm A}_{8(n-1)}{\rm Re_4N_8}}-8E_{\rm A})/8$. Here A represents Na or K atom and $E_{{\rm A}_{8n}{\rm Re_4N_8}}$ is the total energy of ReN$_2$ with the adsorption of $n$ A atom layers. Negative $E_n$ means the adsorption of $n$ layers is accessible and we can obtain the maximum storage capacity by $C_M = xF/M_{\rm ReN_2}$. Here, $F$ is Faraday constant with a value of 26.8 A h mol$^{-1}$, $M_{\rm ReN_2}$ is the molar mass of ReN$_2$ per formula unit, and $x$ means the number of A atoms absorbed on the ReN$_2$ per formula unit.

\begin{figure}[!htbp]
\includegraphics[width=0.4\textwidth]{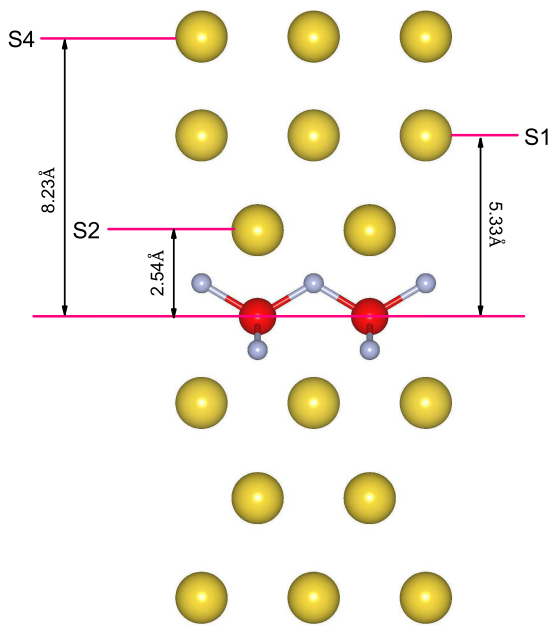}
\caption{\label{fig4}The side view of the atomic structure of Na-intercalated ReN$_2$ monolayer, where up to three Na layers can absorbed on each side.}
\end{figure}

 As shown in FIG.~\ref{fig4}, the ReN$_2$ monolayer can adsorb three layers of Na atoms on each sides for Na-ion batteries, and the maximal storage capacity of the ReN$_2$ monolayer can reach 751 mA h g$^{-1}$. The ReN$_2$ monolayer adsorb only one layer of K atom on each sides and thus its storage capacity is 250 mA h g$^{-1}$ for K-ion batteries. The first Na atom layer is located at the S2 site (above the low-position nitrogen atom), and the average adsorption energy is -1.02 eV, which becomes -0.83 eV for first K atom layer adsorption. Compared to the adsorption energy of one Na/K atom (-1.80/-2.12 eV), this fact shows that the adsorption energy of K atom increases faster than that of Na atom while the concentration of Na/K atoms is increasing. Then for the second layer, Na atoms prefer to stay at the S1 site and the average adsorption energy is -0.20 eV, while the positive average adsorption energy for K atom (0.40 eV) reveals that the second K atom layer fails to be absorbed on the ReN$_2$ monolayer. As for the third layer, Na atoms prefer to be adsorbed at the S4 site and the average adsorption energy becomes -0.08 eV. Three Na atom layers on each sides, in spite of the large mass of the ReN$_2$, make the ReN$_2$ monolayer a high-capacity anode material. The average open-circuit voltage decreases from 1.0 to 0.4 V with the increase of the adsorbed Na concentration from 8 to 24 atoms on the 2$\times$2 supercell. The open circuit voltage for K-ion batteries is 0.83 V with 8 atoms on the 2$\times$2 supercell. In the process of intercalation of Na (K) atom, the change of lattice parameter is only 3.2\% (3.8\%), which is propitious to achieve the rechargeable batteries.

\subsection{Ion diffusion}

\begin{figure}[!htbp]
\includegraphics[width=0.48\textwidth]{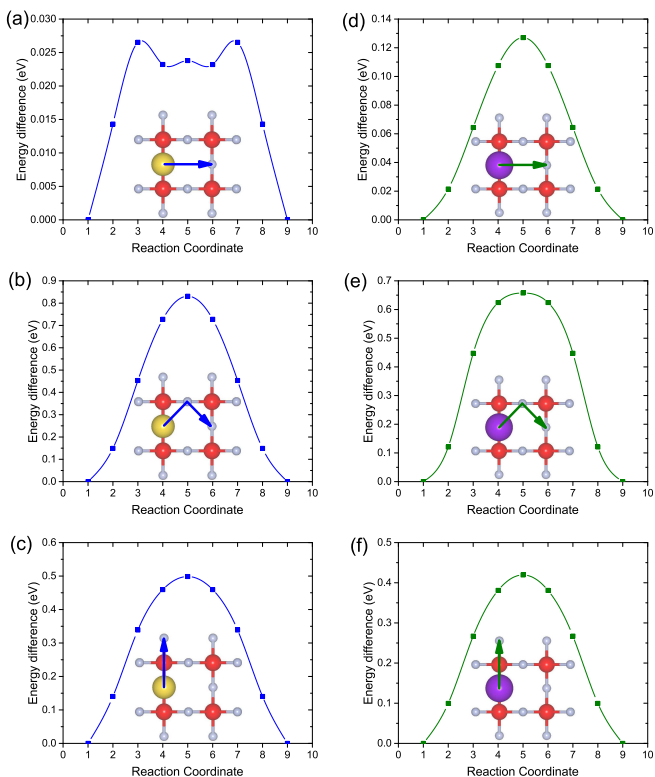}
\caption{\label{fig5}Energy profiles of Na diffusion on path 1 (a), path 2 (b), and path 3 (c), and those of K diffusion on path 1 (d), path 2 (e), and path 3 (f). The inserted figures show the corresponding diffusion pathways.}
\end{figure}

The fast charging and discharging processes need the fast ion diffusion. The ion diffusion depends on the temperature-dependent molecular transition rate $D$, which is proportional to exp($-E_d/k_B$T), where $E_d$, $k_B$, and T are the diffusion barrier, Boltzmann's constant, and temperature, respectively. A low diffusion barrier means a fast charging/discharging process for ion batteries. Three possible diffusion paths are taken into consideration, and the calculated results are shown in FIG.~\ref{fig5}. The diffusion barrier of the path 1 is 0.027 eV and 0.127 eV for Na-ion and K-ion batteries, respectively, which is the lowest of three possible circumstances. It can be explained by the existence of hollow space in the monolayer which reduces the influence of the energy variation at the different sites. The extremely low diffusion barrier of Na atom can bring to ultrafast charging/discharging cycles in the Na-ion batteries.

It is noted that Na/K diffusion on the ReN$_2$ monolayer is quite anisotropic. Coming from the S1 site, Na atom will encounter the hollow space along path1 or relatively high-position Re atom along path 3. It brings to the great anisotropy between the diffusion barriers along the two orthogonal directions, and the ratio is $E_{d3}/E_{d1} \approx $ 18 for Na atom or $E_{d3}/E_{d1} \approx $ 3 for K atom. The anisotropy is greater than phosphorene (the ratio is 8)~\cite{37} and can achieve the unusual transports along the different directions in the these systems.

\subsection{Comparison with others}

It is interesting to compare the ReN$_2$ with other two-dimensional materials for ion batteries. For Na-ion batteries, the data are summarized in TABLE.~\ref{table2}. It can be seen that the maximal storage capacity of the ReN$_2$ monolayer is smaller than those of borophene~\cite{38}, Ca$_2$N~\cite{s14}, phosphorene~\cite{37}, and MoN$_2$ monolayer~\cite{18}, but better than those most of transition-metal dichalcogenides, MXenes, and so on. Actually, the maximum capacity of the ReN$_2$ monolayer is twice or five times that of Sr$_2$N or Mo$_2$C. For K-ion batteries, the maximum capacity of the ReN$_2$ monolayer (250 mA h g$^{-1}$) is comparable to those of GeS (256 mA h g$^{-1}$)~\cite{40}, Mo$_2$C (smaller than 263 mA h g$^{-1}$)~\cite{s11}, and Ti$_3$C$_2$ (191.8 mA h g$^{-1}$)~\cite{s12}, but smaller than those of MoN$_2$ (432 mA h g$^{-1}$)~\cite{18} and BP (570 mA h g$^{-1}$)~\cite{39}. The ReN$_2$ monolayer is better serving as Na-ion storage materials than as K-ion ones. Compared to other two-dimensional materials, the ReN$_2$ have the extremely low Na ion diffusion barrier that is lower than those of other 2D materials except Sr$_2$N monolayer~\cite{s14} and Mo$_2$C monolayer~\cite{s11}. The K-ion diffusion barrier of the ReN$_2$ monolayer (0.127 eV) is smaller than those of MoN$_2$ (0.49 eV)~\cite{18} and BP (0.155 eV)~\cite{39}, and larger than those of Mo$_2$C (0.015 eV)~\cite{s11}, GeS (0.050 eV)~\cite{40}, and Ti$_3$C$_2$ (0.103 eV)~\cite{s12}. Very importantly, the ReN$_2$ monolayer have both very high storage capacity and extremely low ion diffusion barrier, which makes it promising electrode materials for Na-ion batteries.

\begin{table}[!htbp]
\caption{\label{table2} Comparison of maximum capacity $C_M$ and diffusion barrier $E_d$ of 12 two-dimensional electrode materials for Na-ion batteries.}
\begin{ruledtabular}
\begin{tabular}{cccc}
 & $C_M$ (mA h g$^{-1}$)& $E_d$ (eV) & Ref. \\ \hline
$\beta _{12}$ borophene  & 1984 & 0.33 & Ref.~\cite{38}\\
$\chi _3$ borophene    & 1240 & 0.34 & Ref.~\cite{38}\\
Ca$_2$N & 1138 & 0.08 & Ref.~\cite{s14}\\
Phosphorene & 865 & 0.04 & Ref.~\cite{37}\\
MoN$_2$ & 864 & 0.56 & Ref.~\cite{18}\\
GeS & 512 & 0.09 & Ref.~\cite{40}\\
TiS$_2$ & 339 & 0.22 & Ref.~\cite{41}\\
Sr$_2$N & 283 & 0.016 & Ref.~\cite{s14}\\
NbS$_2$ & 263 & 0.07 & Ref.~\cite{41}\\
BP & 143 & 0.217 & Ref.~\cite{39}\\
Mo$_2$C & 132 & 0.019 & Ref.~\cite{s11}\\
W$_2$C  & 113 & 0.019 & Ref.~\cite{added}\\
ReN$_2$                   & 751 & 0.027 & This work\\
\end{tabular}
\end{ruledtabular}
\end{table}

\section{Conclusion}

In summary, we have proposed the ReN$_2$ monolayer as new 2D materials and superior anode materials for Na-ion and K-ion batteries by first-principles investigation. Our calculated results show that the two-dimensional ReN$_2$ monolayer is promising as electrode materials because (1) it keeps metallic feature before and after the adsorption of Na/K atom and thus has the good electric conductivity; (2) the small lattice changes during the intercalation reveals the good recyclability; (3) the maximum storage capacity of the ReN$_2$ monolayer reaches to 751 mA h g$^{-1}$ for Na-ion batteries, which is quite high among two-dimensional electrode materials to date; (4) the low ion diffusion barriers (0.027 eV for Na and 0.127 eV for K) can make the charging/discharging cycles fast. Therefore, our first-principles investigation shows that the ReN$_2$ monolayer is superior as two-dimensional electrode materials for Na/K-ion batteries.

\begin{acknowledgments}
This work is supported by the Nature Science Foundation of China (Grant No. 11574366), by the Strategic Priority Research Program of the Chinese Academy of Sciences (Grant No.XDB07000000), and by the Department of Science and Technology of China (Grant No. 2016YFA0300701).
The calculations were performed in the Milky Way \#2 supercomputer system at the National Supercomputer Center of Guangzhou, Guangzhou, China.
\end{acknowledgments}


\section*{References}

\begin{thebibliography}{23}

\bibitem{1} K. S. Novoselov, A. K. Geim, S. V. Morozov, D. Jiang, Y. Zhang, S. V. Dubonos, I. V. Grigorieva, A. A. Firsov, Electric Field Effect in Atomically Thin Carbon Films, Science 306, 666 (2004).

\bibitem{2} K. S. Novoselov, A. K. Geim, S. V. Morozov, D. Jiang, M. I. Katsnelson, I. V. Grigorieva, S. V. Dubonos, A. A. Firsov, Two-dimensional gas of massless Dirac fermions in graphene, Nature 438, 197 (2005).

\bibitem{3} A. K. Geim, K. S. Novoselov, The rise of graphene, Nat. Mater. 6, 183 (2007).

\bibitem{4} A. H. Castro Neto, F. Guinea, N. M. R. Peres, K. S. Novoselov, A. K. Geim, The electronic properties of graphene, Rev. Mod. Phys. 81, 109 (2009).

\bibitem{5} B. Radisavljevic, A. Radenovic, J. Brivio, V. Giacometti, A. Kis, Single-layer MoS$_2$ transistors, Nat. Nanotech. 6, 147 (2011).

\bibitem{6} L. Li, Y. Yu, G. J. Ye, Q. Ge, X. Ou, H. Wu, D. Feng, X. H. Chen, Y. Zhang, Black phosphorus field-effect transistors, Nat. Nanotech. 9, 372 (2014).

\bibitem{7} Z. Yin, H. Li, H. Li, L. Jiang, Y. Shi, Y. Sun, G. Lu, Q. Zhang, X. Chen, H. Zhang, Single-Layer MoS$_2$ Phototransistors, ACS Nano 6, 74 (2012).

\bibitem{8} K. Xu, Z. Wang, F. Wang, Y. Huang, F. Wang, L. Yin, C. Jiang, J. He, Ultrasensitive Phototransistors Based on Few-Layered HfS$_2$, Adv. Mater. 27, 7881 (2015).

\bibitem{9} M. Y. Li, Y. Shi, C. C. Cheng, L. S. Lu, Y. C. Lin, H. L. Tang, M. L. Tsai, C. W. Chu, K. H. Wei, J. H. He, Epitaxial growth of a monolayer WSe$_2$-MoS$_2$ lateral p-n junction with an atomically sharp interface, Science 349, 524 (2015).

\bibitem{10} B. Peng, G. Yu, X. Liu, B. Liu, X. Liang, L. Bi, L. Deng, T. C. Sum, K. P. Loh, Ultrafast charge transfer in MoS$_2$/WSe$_2$ p-n Heterojunction, 2D Mater. 3, 025020 (2016).

\bibitem{11} M. Acerce, D. Voiry, M. Chhowalla, Metallic 1T phase MoS$_2$ nanosheets as supercapacitor electrode materials, Nat. Nanotech. 10, 313 (2015).

\bibitem{12} H. Yang, N. Wang, Q. Xu, Z. Chen, Y. Ren, J. M. Razal, J. Chen, Fabrication of graphene foam supported carbon
nanotube/polyaniline hybrids for high-performance supercapacitor applications, 2D Mater. 1, 034002 (2014).

\bibitem{13} B. Dunn, H. Kamath, J. M. Tarascon, Electrical Energy Storage for the Grid: A Battery of Choices, Science 334, 928   (2011).

\bibitem{14} E. Quesnel, F. Roux, F. Emieux, P. Faucherand, E. Kymakis, G. Volonakis, F. Giustino, B. Martin-Garcia, I. Moreels, S. A. Gursel, Graphene-based technologies for energy applications, challenges and perspectives, 2D Mater. 2, 030204  (2015).

\bibitem{l1} J. Tarascon, M. Armand, Issues and challenges facing rechargeable lithium batteries, Nature 414, 359 (2001).

\bibitem{l2} Y. Idota, T. Kubota, A. Matsufuji, Y. Maekawa, T. Miyasaka, Tin-Based Amorphous Oxide: A High-Capacity Lithium-Ion-Storage Material, Science 276, 1395 (1997).

\bibitem{l3} K. T. Nam, D. W. Kim, P. J. Yoo, C. Y. Chiang, N. Meethong, P. T. Hammond, Y. M. Chiang, A. M. Belcher, Virus-enabled synthesis and assembly of nanowires for lithium ion battery electrodes, Science 312, 885 (2006).

\bibitem{l4} F. Zou, X. Hu, L. Qie, Y. Jiang, X. Xiong, Y. Qiao, Y. Huang, Facile synthesis of sandwiched Zn$_2$GeO$_4$-graphene oxide nanocomposite as a stable and high-capacity anode for lithium-ion batteries, Nanoscale 6, 924 (2014).

\bibitem{l6} X. Sun, C. Yan, Y. Chen, W. Si, J. Deng, S. Oswald, L. Liu, O. G. Schmidt, Three-dimensionally "curved" NiO nanomembranes as ultrahigh rate capability anodes for Li-ion batteries with long cycle lifetimes, Adv. Energy Mater. 4, 1300912 (2014).

\bibitem{l7} J. Wang, Q. Zhang, X. Li, B. Zhang, L. Mai, K. Zhang, Smart construction of three-dimensional hierarchical tubular transition metal oxide core/shell heterostructures with high-capacity and long-cycle-life lithium storage, Nano Energy 12, 437 (2015).

\bibitem{l8} X. Qian, X. Gu, M. S. Dresselhaus, R. Yang, Anisotropic Tuning of Graphite Thermal Conductivity by Lithium Intercalation, J. Phys. Chem. Lett. 7, 4744 (2016).

\bibitem{l9} L. Shi, T. Zhao, A. Xu, J. Xu, Ab initio prediction of a silicene and graphene heterostructure as an anode material for Li- and Na-ion batteries, J. Mater. Chem. A 4, 16377 (2016).

\bibitem{s1} V. Etacheri, R. Marom, R. Elazari, G. Salitra, D. Aurbach, Challenges in the development of advanced Li-ion batteries: A review, Energy Environ. Sci. 4, 3243 (2011).

\bibitem{s2} M. Reddy, G. Subba Rao, B. Chowdari, Metal oxides and oxysalts as anode materials for Li ion batteries, Chem. Rev. 113, 5364 (2013).

\bibitem{s3} M. T. McDowell, S. W. Lee, W. D. Nix, Y. Cui, 25th anniversary article: Understanding the lithiation of silicon and other alloying anodes for lithium-ion batteries, Adv. Mater. 25, 4966 (2013).

\bibitem{s4} D. Larcher, J. Tarascon, Towards greener and more sustainable batteries for electrical energy storage, Nat. Chem. 7, 19 (2015).

\bibitem{s6} E. Pollak, B. Geng, K.-J. Jeon, I. T. Lucas, T. J. Richardson, F. Wang, R. Kostecki, The interaction of Li+ with single-layer and few-layer graphene, Nano Lett. 10, 3386 (2010).

\bibitem{s7} Q. Tang, Z. Zhou, Z. Chen, Graphene-related nanomaterials: Tuning properties by functionalization, Nanoscale 5, 4541 (2013).

\bibitem{s8} L. David, R. Bhandavat, G. Singh, MoS$_2$/graphene composite paper for sodium-ion battery electrodes, ACS Nano 8, 1759 (2014).

\bibitem{s9} R. Bhandavat, L. David, G. Singh, Synthesis of surface-functionalized WS$_2$ nanosheets and performance as li-ion battery anodes, J. Phys. Chem. Lett. 3, 1523 (2012).

\bibitem{s10} Y. Jing, Z. Zhou, C. Cabrera, Z. Chen, Metallic VS$_2$ monolayer: A promising 2D anode material for lithium ion batteries, J. Phys. Chem. C 117, 25409 (2013).

\bibitem{s11} Q. Sun, Y. Dai, Y. Ma, T. Jing, W. Wei, B. Huang, Ab Initio Prediction and Characterization of Mo$_2$C Monolayer as Anodes for Lithium-Ion and Sodium-Ion Batteries, J. Phys. Chem. Lett. 7, 937 (2016).

\bibitem{s12} D. Er, J. Li, M. Naguib, Y. Gogotsi, V. Shenoy, Ti$_3$C$_2$ MXene as a high capacity electrode material for metal (Li, Na, K, Ca) ion batteries, ACS Appl. Mater. Interfaces 6, 11173 (2014).

\bibitem{s13} J. Hu, B. Xu, C. Ouyang, S. A. Yang, Y. Yao, Investigations on V$_2$C and V$_2$CX$_2$ (X = F, OH) monolayer as a promising anode material for Li ion batteries from first-principles calculations, J. Phys. Chem. C 118, 24274 (2014).

\bibitem{added} A. Samad, A. Shafique, H. J. Kim, and Y.-H. Shin, Superionic and electronic conductivity in momlayer W$_2$C: ab initio predictions, J. Mater. Chem. A 5, 11094 (2017).

\bibitem{s14} J. Hu, B. Xu, S. A. Yang, S. Guan, C. Ouyang, Y. Yao, 2D Electrides as Promising Anode Materials for Na-Ion Batteries from First-Principles Study, ACS Appl. Mater. Interfaces 7, 24016 (2015).


\bibitem{22} P. E. Blochl, Projector Augmented-Wave Method, Phys. Rev. B 50, 17953 (1994).

\bibitem{23} G. Kresse, J. Furthmller, Efficient iterative schemes for ab initio total-energy calculations using a plane-wave basis set, Phys. Rev. B 54, 11169 (1996).

\bibitem{24} J. P. Perdew, K. Burke, M. Ernzerhof, Generalized Gradient Approximation Made Simple, Phys. Rev. Lett. 77, 3865 (1996).

\bibitem{25} H. J. Monkhorst, J. D. Pack, Special points for Brillouin-zone integrations, Phys. Rev. B 13, 5188 (1976).

\bibitem{29} A. Togo, F. Oba, I. Tanaka, First-principles calculations of the ferroelastic transition between rutile-type and CaCl$_2$-type SiO$_2$ at high pressures, Phys. Rev. B 78, 134106 (2008).


\bibitem{26} J. Heyd, G. E. Scuseria, M. Ernzerhof, Hybrid functionals based on a screened Coulomb potential, J. Chem. Phys. 118, 8207 (2003).

\bibitem{27} A. V. Krukau, O. A. Vydrov, A. F. Izmaylov, G. E. Scuseria, Influence of the exchange screening parameter on the performance of screened hybrid functionals, J. Chem. Phys. 125, 224106 (2006).

\bibitem{28} P. M. Sanchez, A. J. Cohen, W. Yang, Localization and Delocalization Errors in Density Functional Theory and Implications for Band-Gap Prediction, Phys. Rev. Lett. 100, 146401 (2008).

\bibitem{D2} S. Grimme, Semiempirical GGA-type density functional constructed with a long-range dispersion correction, J. Comput. Chem. 27, 1787 (2006).

\bibitem{19} F. Kawamura, H. Yusa, T. Taniguchi, Synthesis of rhenium nitride crystals with MoS$_2$ structure, Appl. Phys. Lett. 100, 251910 (2012).

\bibitem{20} Y. Wang, T. Yao, J. L. Yao, J. Zhang, H. Gou, Does the real ReN$_2$ have the MoS$_2$ structure? Phys. Chem. Chem. Phys. 15, 183 (2013).

\bibitem{21} H. Yan, M. Zhang, Q. Wei, P. Guo, Theoretical study on tetragonal transition metal dinitrides from first
principles calculations, J. Alloys Compounds 581, 508 (2013).


\bibitem{30} M. Chhowalla, H. S. Shin, G. Eda, L. J. Li, K. P. Loh, H. Zhang, The chemistry of two-dimensional layered
transition metal dichalcogenide nanosheets, Nat. Chem. 5, 263 (2013).

\bibitem{31} M. Calandra, Chemically exfoliated single-layer MoS$_2$: Stability, lattice dynamics, and catalytic adsorption
from first principles, Phys. Rev. B 88, 245428 (2013).

\bibitem{32} S. Tongay, H. Sahin, C. Ko, A. Luce, W. Fan, K. Liu, J. Zhou, Y. S. Huang, C. H. Ho, J. Yan, D. F. Ogletree, S. Aloni, J. Ji, S. Li, J. Li, F. M. Peeters, J. Wu, Monolayer behaviour in bulk ReS$_2$ due to electronic
and vibrational decoupling, Nat. Commun. 5, 3252 (2014).

\bibitem{33} D. H. Keum, S. Cho, J. H. Kim, D. H. Choe, H. J. Sung, M. Kan, H. Kang, J. Y. Hwang, S. W. Kim, H. Yang, K. J. Chang, Y. H. Lee, Band gap opening in few-layered monoclinic MoTe$_2$, Nat. Phys. 11, 482 (2015).

\bibitem{34} F. Ersan, S. Cahangirov, G. Gokoglu, A. Rubio, E. Akturk, Stable monolayer honeycomb-like structures of RuX$_2$ (X = S, Se), Phys. Rev. B 94, 155415 (2016).

\bibitem{mon2ce} H. Wu, Y. Qiana, R. Luo, W. Tan, A theoretical study on the electronic property of a new two-dimensional material molybdenum dinitride, Phys. Lett. A 380 768 (2016).

\bibitem{35} F. Liu, P. Ming, J. Li, Ab initio calculation of ideal strength and phonon instability of graphene under tension, Phys. Rev. B 76, 064120 (2007).

\bibitem{prb} R. C. Andrew, R. E. Mapasha, A. M. Ukpong, N. Chetty, Mechanical properties of graphene and boronitrene, Phys. Rev. B 85, 125428 (2012).

\bibitem{36} S. Zhao, Z. Li, J. Yang, Obtaining Two-Dimensional Electron Gas in Free Space without
Resorting to Electron Doping: An Electride Based Design, J. Am. Chem. Soc. 136, 13313 (2014).

\bibitem{37} V. V. Kulish, O. I. Malyi, C. Persson, Ping Wu, Phosphorene as an anode material for Na-ion
batteries: a first-principles study, Phys. Chem. Chem. Phys. 17, 13921 (2015).

\bibitem{38} X. Zhang, J. Hu, Y. Cheng, H. Y. Yang, Y. Yao, S. A. Yang, Borophene as an extremely high capacity
electrode material for Li-ion and Na-ion batteries, Nanoscale 8, 15340 (2016).

\bibitem{18} X. Zhang, Z. Yu, S. S. Wang, S. Guan, H. Y. Yang, Y. Yao, S. A. Yang, Theoretical prediction of MoN$_2$ monolayer as a high capacity electrode material for metal ion batteries, J. Mater. Chem. A 4, 15224 (2016).

\bibitem{39} H. R. Jiang, W. Shyy, M. Liu, L. Wei, M. C. Wu, T. S. Zhao, Boron phosphide monolayer as a potential anode
material for alkali metal-based batteries, J. Mater. Chem. A 5 672 (2017).

\bibitem{40} F. Li, Y. Qu, M. Zhao, Germanium sulfide nanosheet: a universal anode material for alkali metal ion batteries, J. Mater. Chem. A 4, 8905 (2016).

\bibitem{41} E. Yang, H. Ji, Y. Jung, Two-Dimensional Transition Metal Dichalcogenide Monolayers as Promising Sodium Ion Battery Anodes, J. Phys. Chem. C 119, 26374 (2015).

\end{thebibliography}
\end{document}